\documentclass[a4paper,12pt]{article}
\usepackage[dvips]{graphicx}
\usepackage{amssymb}
\usepackage{amscd}
\usepackage{mathrsfs}
\usepackage{latexsym}
\usepackage{makeidx}
\date{}
\author{Claudio Garola\footnote{E-mail address: garola@le.infn.it}, \quad Sandro Sozzo\footnote{E-mail address: sozzo@le.infn.it} \\ Dipartimento di Fisica and Sezione INFN \\Universit\`a di Lecce, via Arnesano, 73100 LECCE  \vspace{.5cm} \\ and \vspace{.5cm} \\
Jaros\l aw Pykacz\footnote{E-mail address: pykacz@math.univ.gda.pl} \\
Instytut Matematyki, Uniwersytet Gda\'{n}ski \\ 80-952, Gda\'{n}sk, Poland}
\title{\textbf{Quantum Machine and SR Approach: a Unified Model}}
\begin{document}
\maketitle
\begin{abstract}
\noindent
The Geneva--Brussels approach to quantum mechanics (QM) and the semantic realism (SR) nonstandard interpretation of QM exhibit some common features and some deep conceptual differences. We discuss in this paper two elementary models provided in the two approaches as intuitive supports to general reasonings and as a proof of consistency of general assumptions, and show that Aerts' \emph{quantum machine} can be embodied into a macroscopic version of the \emph{microscopic SR model}, overcoming the seeming incompatibility between the two models. This result provides some hints for the construction of a unified perspective in which the two approaches can be properly placed.

\vspace{.75cm}
\noindent 
\textbf{Key Words}. quantum mechanics; quantum machine; semantic realism. 
\end{abstract}
\section{Introduction}
The \emph{Geneva--Brussels} (\emph{GB}) \emph{approach} to quantum mechanics (QM) is well known. It was started by Jauch and Piron in Geneva \cite{j68}, \cite{p76} and then continued by Aerts and his collaborators in Brussels \cite{a82}--\cite{a04}. It can be classified in the field of ``quantum structures research'' \cite{a99b}, aiming both at basing QM on fundamental concepts that can be operationally defined and at providing a physical justification of the relations established by QM among these concepts. In its latest version it also proposes, however, some fundamental changes of the standard theory in order to avoid a number of quantum problems and paradoxes and overcome the limits of the QM description of the physical world (with special attention to compound quantum systems).

The GB approach exhibits two relevant features. Firstly, quantum probabilities are interpreted as \emph{epistemic} (they express our ignorance about \emph{hidden measurements} rather than about hidden states of the physical system), at variance with the standard interpretation, where quantum probabilities are (mostly) ontologic. Secondly, the mathematical structure of the set of \emph{empirical propositions} called quantum logic (QL) is interpreted as a consequence of our ``possibilities of active experimenting'' on physical systems, not as a new logic formalizing some ``process of our reflection'', so that QL does not characterize the microscopic world (indeed, when our possibilities of active experimenting on a \emph{macroscopic} entity are suitably limited, one can find quantum logical structures associated with this entity).

The above features also appear within the \emph{semantic realism} (\emph{SR}) \emph{interpretation} of QM propounded by the Lecce research group on the foundations of QM as an alternative to the standard interpretation \cite{gs96a}--\cite{ga02}, aiming to avoid the same problems and paradoxes considered by the GB approach. Moreover, the SR interpretation also implies that a broader theory embodying QM is, at least in principle, possible. This suggests that a comparison between the two perspectives could be interesting, and that an attempt at extablishing links between them could be fruitful. Yet, whenever one starts this job, one immediately meets a serious difficulty. Indeed, the GB approach is highly contextual, following in this sense the standard QM tradition. On the contrary, the SR interpretation mantains that contextuality is the root of most quantum paradoxes and elaborates a strategy (based on some epistemological criticisms to the standard interpretation) to avoid it without conflicting with the mathematical apparatus and the predictions of QM. Thus the two approaches seem conceptually incompatible at first sight. However, a deeper insight shows that this is not necessarily the case. We cannot yet prove this by providing a general perspective in which both approaches find a proper place, but we can show that an integration is possible in the case of the models introduced in the GB and SR approaches as intuitive supports to general reasonings (and also as a demonstration of consistency of some abstract assumptions, especially within the SR interpretation). To be precise, we intend to show in the present paper that a macroscopic version of the intuitive picture for the SR model provided in some previous papers \cite{ga03}, \cite{gp04} (briefly, \emph{microscopic SR model} in the following) can be constructed which embodies Aerts' \emph{quantum machine} (which plays an important role in Aerts' approach since it provides a macroscopic model for spin measurements on spin--$\frac{1}{2}$ quantum systems). This \emph{unified} (\emph{SR}) \emph{model} provides the same predictions as Aerts' quantum machine whenever one takes into account those and only those samples of the physical system under investigation that are actually detected if a measurement is performed. In this sense we can say that the two models are formally equivalent (the equivalence is attained, however, by means of a rather artificial and complicate construction, which does not aim to represent any physical reality but only to illustrate a logical possibility).

It must still be stressed that our unified model applies to quantum systems described by two dimensional Hilbert spaces, just as the quantum machine. The GB approach provides, however, more general models which apply to higher dimensional quantum cases. Embodying these models within a generalized unified (SR) model seems possible in principle (the microscopic SR model makes no reference to the dimensionality of the Hilbert space of the system) but it may raise some problems. In particular, it could be difficult in this case to reconciliate the contextuality of the GB models with the noncontextuality of the SR model. We do not discuss this problem in the present paper and limit ourselves to note that the remarks in Sec. 4.1 on the different notions of contextuality introduced in the literature may help in solving it.

Finally, let us briefly resume the content of the various sections of our paper. Firstly, we sketch the guidelines of the microscopic SR model and quantum machine in Secs. 2 and 3, respectively. This leads us, in particular, to complete the microscopic SR model by means of some equations which do not appear in the original draft. Then, our unified model is introduced in Sec. 4, it is discussed in the case of pure states in Sec. 4.1, and it is generalized to the case of mixed states in Sec. 4.2.

\section{The microscopic SR model}
As we have anticipated in the Introduction, the consistency of the SR approach has recently been demonstrated by means of a set--theoretical model, the microscopic SR model, that shows, circumventing known no--go theorems, how a local and noncontextual (hence objective) picture of the microworld can be constructed without altering the formalism and the (statistical) interpretation of QM. We report the essentials of it here.

To begin with, let us accept the standard notion of state of a physical system $\Omega$ as a class of physically equivalent preparing devices \cite{l83}. Furthermore, let us call \emph{physical object} any individual sample $x$ of $\Omega$ obtained by activating a preparing device, and say that \emph{x is in the state $S$} if the device $\pi $ preparing $x$ belongs to $S$. Whenever $\Omega$ is a microscopic physical system, let us introduce a set $\mathcal{E}$ of \emph{microscopic} physical properties that characterize $\Omega$ and play the role of theoretical entities. For every physical object $x$, every property $f \in \mathcal{E}$ is associated with $x$ in a dichotomic way, so that one briefly says that every $f \in \mathcal{E}$ either is possessed or it is not possessed by $x$. This is the main difference between the SR interpretation and the ortodox interpretation of QM, in which it is assumed that microphysical objects generally do not possess a property until it is measured \cite{me93}. The set $\mathcal{F}_{0}$ of all \emph{macroscopic} properties is then introduced as in standard QM, that is, it is defined as the set of all pairs of the form $({\mathcal A}_{0},\Delta)$, where ${\mathcal A}_{0}$ is an observable (that is, a class of physically equivalent measuring apparatuses) with spectrum $\Lambda _{0}$, and $\Delta $ a Borel set on the real line $\Re$ (for every observable ${\mathcal A}_{0}$, different Borel sets containing the same subset of $\Lambda _{0}$ obviously define physically equivalent properties; we note explicitly that, whenever we speak about macroscopic properties in the following, we actually understand such classes of physically equivalent macroscopic properties). Yet, every observable ${\mathcal A}_{0}$ is obtained from a suitable observable $\mathcal{A}$ of standard QM by adding to the spectrum $\Lambda $ of $\mathcal{A}$ a further outcome $a_{0}$ that does not belong to $\Lambda$, called the \emph{no--registration} outcome of $\mathcal{A}_{0}$ (note that such an outcome is introduced also within the standard quantum theory of measurement, but it plays here a different theoretical role), so that $\Lambda _{0}=\Lambda\cup \left\{ a_{0} \right\}$. The set $\mathcal{E}$ of all microscopic properties is then assumed to be in one--to--one correspondence with the subset ${\mathcal F} \subseteq {\mathcal F}_{0}$\ of all macroscopic properties of the form $F=({\mathcal A}_{0},\Delta)$, where ${\mathcal A}_{0}$ is an observable and $a_{0}\notin \Delta$.

Basing on the above definitions and assumptions, one can provide the following description of the measurement process. Whenever a physical object $x$ is prepared in a state $S$ by a given device $\pi $, and ${\mathcal A}_{0}$ is measured by means of a suitable apparatus, the set of microscopic properties possessed by $x$ produces a probability (which is either 0 or 1 if the model is \emph{deterministic}) that the apparatus does not react, so that the outcome $a_{0}$ may be obtained. In this case, $x$ is not detected and one cannot get any explicit information about the microscopic physical properties possessed by $x$. If, on the contrary, the apparatus reacts, an outcome different from $a_{0}$, say $a$, is obtained, and one is informed that $x$ possesses all microscopic properties associated with macroscopic properties of the form $F=({\mathcal A}_{0},\Delta)$, where $\Delta $ is a Borel set such that $a_{0}\notin \Delta $ and $a \in \Delta $ (for the
sake of brevity we also say that $x$ \textit{possesses} all macroscopic properties as $F$ in this case).

In order to place properly quantum probability within the above picture, let us consider a preparing device $\pi \in S$ that is activated repeatedly. In this case a (finite) set $\mathscr{S}$ of physical objects in the state $S$ is prepared. Then, let us partition $\mathscr{S}$ into subsets ${\mathscr S}^{1}$, ${\mathscr S}^{2}$, ..., ${\mathscr S}^{n}$, such that in each subset all objects possess the same \emph{microscopic} properties (we can briefly say that the objects in ${\mathscr S}^{i}$, possessing the same microscopic properties, are in the same \emph{microstate} $S^{i}$), and assume that a measurement of an observable ${\mathcal A}_{0}$ is done on every object. Finally, let us introduce the following symbols (see \textbf{Fig. 1}).

\vspace{1cm}
\begin{center}
\begin{tabular}{|c|c|c|c|c|c|c|}
\hline
${\mathscr S}^{1}$ & \ldots  & ${\mathscr S}^{i}$ & \ldots   & ${\mathscr S}^{j}$ &  \ldots   & ${\mathscr S}^{n}$  \\ 
$(g,f,h,..)^{1}$ & \ldots &  $(g,f,\neg h..)^{i}$ &   \ldots  & $(g,\neg f,h,..)^{j}$ & \ldots & $(\neg g,\neg f,h,..)^{n}$  \\
$N^{1}$ & \ldots  & $N^{i}$  &   \ldots    & $N^{j}$ &  \ldots   & $N^{n}$  \\ 
\hline
\hline
 & & & & &  & \\
 & & & & & &  \\ 
 & & & & & &  \\ 
 & & & & & &  \\ 
 & & & & & &  \\ 
 & & & & & &  \\ 
$N_F^{1}=$ & & $N_{F}^{i}=$ & & $N_F^{j}=0$ & & $N_F^{n}=0$  \\
$N^{1}-N_0^{1}$ & & $N^{i}-N_0^{i}$  & & & &  \\ 
 & & & & & &  \\ 
 & & & & & &  \\ 
\cline{6-6} & & & & & & \\ 
\cline{3-3} & & & & & & \\
\cline{1-1} & & & & & & \\
\cline{7-7} & & & & & & \\
\cline{5-5} & & & & & & \\
\cline{4-4} \cline{2-2} & & & & & & \\
$N_0^{1}$ & &  $N_0^{i}$  & &  $N_0^{j}$ &  & $N_0^{n}$  \\
 & & & & & &  \\ 
\hline
\end{tabular}
\end{center}

\vspace{.75cm}
\noindent
\textbf{Fig. 1}. \emph{Set--theoretical representation of the general SR model. The property $F$ is the macroscopic property corresponding to the microscopic property $f$.}

\vspace{1cm}
(i) The number $N$ of physical objects in ${\mathscr S}$.

(ii) The number $N_{0}$ of physical objects in ${\mathscr S}$ that are not detected.

(iii) The number $N^{i}$ of physical objects in ${\mathscr S}^{i}$.

(iv) The number $N_{0}^{i}$ of physical objects in ${\mathscr S}^{i}$ that are
not detected.

(v) The number $N_{F}^{i}$  of physical objects in ${\mathscr S}^{i}$ that possess the macroscopic property $F=({\mathcal A}_{0},\Delta)$, with $a_0 \notin \Delta$, corresponding to the microscopic property $f$.

It is apparent that the number $N_{F}^{i}$ either coincides with $N^{i}-N_{0}^{i}$ or with $0$. The former case occurs whenever $f$ is possessed by the objects in ${\mathscr S}^{i}$, since all objects that are detected then yield outcome in $\Delta$. The latter case occurs whenever $f$ is not possessed by the objects in ${\mathscr S}^{i}$, since all objects that are detected then yield outcome different from $a_{0}$ but outside $\Delta $ (note that the microscopic property $\neg f$ corresponding to $F^{\perp}=({\mathcal A}_{0}, \Re \setminus (\Delta \cup \{ a_0 \}))$ is possessed by the objects in  ${\mathscr S}^{i}$ in this case). In both cases one can assume that $N^{i}-N_{0}^{i}\neq 0$\footnote{Note that in a deterministic model either $N_0^i=0$ or $N_0^i=N^i$, hence either $N^i-N_0^{i}=N^{i}$ or $N^i-N_0^{i}=0$, so that the assumption $N^{i}-N_{0}^{i}\neq 0$ does not hold. However, $N^{i}-N_{0}^{i}=0$ implies $N_F^{i}=0$, and eq. (\ref{macro2}) can be recovered by modifying our reasonings in an obvious way.}, so that the following equation holds.
\begin{equation} \label{micro1}
\frac{N_{F}^{i}}{N^{i}}=\frac{N^{i}-N_{0}^{i}}{N^{i}}\frac{N_{F}^{i}}{N^{i}-N_{0}^{i}}.
\end{equation}
The term on the left in eq. (\ref{micro1}) represents the fraction of objects possessing the property $F$ in $\mathscr{S}^{i}$, the first term on the right the fraction of objects in ${\mathscr S}^{i}$ that are detected, the second term (which is either $0$ or $1$) indicates whether the objects in ${\mathscr S}^{i}$ that are detected possess the property $F$ or not.

The fraction of objects in ${\mathscr S}$ that possess the property $F$ is given by
\begin{equation} \label{macro1}
\frac{1}{N}\sum_{i}N_{F}^{i}=\frac{N-N_{0}}{N} \Big (\sum_{i}\frac{N_{F}^{i}}{N-N_{0}} \Big ).
\end{equation}
Let us assume now that all fractions of objects converge in the large number limit, so that they can be substituted by probabilities, and that these probabilities do not depend on the choice of the preparing device $\pi$ in $S$. Hence, if one considers the large number limit of eq. (\ref{micro1}), one gets
\begin{equation} \label{micro2}
{\mathscr P}_{S}^{i,t}(F)={\mathscr P}_{S}^{i,d}(F) {\mathscr P}_{S}^{i}(F),
\end{equation}
where ${\mathscr P}_{S}^{i,t}(F)$ is the total probability that a physical object $x$ which possesses the microscopic properties that characterize ${\mathscr S}^{i}$, \emph{i.e.}, which is in the state $S^{i}$, also possesses the property $F$, ${\mathscr P}_{S}^{i,d}(F)$ is the probability that $x$ is detected when $F$ is measured on it, ${\mathscr P}_{S}^{i}(F)$ (which is either $0$ or $1$) is the probability that $x$ possesses the property $F$ when detected. Analogously, the large number limit of eq. (\ref{macro1}) yields
\begin{equation} \label{macro2}
{\mathscr P}_{S}^{t}(F)={\mathscr P}_{S}^{d}(F) {\mathscr P}_{S}(F),
\end{equation}
where ${\mathscr P}_{S}^{t}(F)$ is the total probability that a physical object $x$ in a state $S$ possesses the property $F$, ${\mathscr P}_{S}^{d}(F)$ is the probability that $x$ is detected when $F$ is measured on it, ${\mathscr P}_{S}(F)$ is the probability that $x$ possesses the property $F$ when detected.

If we identify the previous probabilities with the corresponding fractions of objects in the large number limit, it is possible to express the macroscopic probabilities in eq. (\ref{macro2}) in terms of the microscopic probabilities in eq. (\ref{micro2}). Indeed,
\begin{displaymath}
{\mathscr P}_{S}^{d}(F)=\frac{N-N_0}{N}=\frac{1}{N}\sum_{i} (N^{i}-N_{0}^{i})=\sum_{i} \Big ( \frac{N^{i}}{N}\frac{N^{i}-N_{0}^{i}}{N^{i}} \Big )=
\end{displaymath}
\begin{equation} \label{newprobability-d}
=\sum_{i} {\mathscr P}(S^{i}|S) {\mathscr P}_{S}^{i,d}(F),
\end{equation}
where we have identified the fraction of objects in the microstate $S^{i}$ with respect to the objects in the state $S$ with the conditional probability ${\mathscr P}(S^{i}|S)$ that an object $x$ in the state $S$ actually is in the microstate $S^{i}$. Analogously, we get
\begin{displaymath} 
{\mathscr P}_{S}^{t}(F)=\frac{1}{N}\sum_{i}N_{F}^{i}=\sum_{i}\Big ( \frac{N^{i}}{N}\frac{N^i-N_0^i}{N^{i}}\frac{N_{F}^{i}}{N^i-N_0^i} \Big )=
\end{displaymath}
\begin{equation} \label{newprobability-t}
=\sum_{i} {\mathscr P}(S^{i}|S) {\mathscr P}_{S}^{i,d}(F){\mathscr P}_{S}^{i}(F).
\end{equation}

The interpretation of ${\mathscr P}_{S}(F)$ makes it reasonable to identify this probability with the quantum probability that a physical object in the state $S$ possesses the property $F$. Hence, standard QM can be recovered within the model as the theory that allows one to evaluate ${\mathscr P}_{S}(F)$ (and its evolution in time) for every system $\Omega$, state $S$, and property $F=({\mathcal A}_{0},\Delta)$ such that $a_{0}\notin \Delta $. In this perspective, no change of the formalism and the statistical interpretation of standard QM is required. In particular, any state $S$ can be represented, as usual, by means of a trace class operator $\rho _{S}$ on a Hilbert space $\mathscr {H}$ associated with $\Omega$ and any macroscopic property $F$ that corresponds to a microscopic property can be represented by means of a projection operator $P_{F}$ on $\mathscr{H}$, so that ${\mathscr P}_{S}(F)=Tr \{\rho_{S}P_{F} \}$. Thus, the model provides a picture of the microworld which embodies standard QM. This picture is objective, in the sense that for every physical object $x$ in the state $S$, every macroscopic property of the form $F=(\mathcal{A}_{0},\Delta)$ (where $a_{0}$ may now belong or not to $\Delta$) either is possessed or is not possessed by $x$, and the probability that it is possessed/not possessed is determined by the microscopic properties possessed by $x$, which do not depend on the measuring apparatus (hence microscopic properties play in the model a role similar to states in objective local theories \cite{ch74}). 

Objectivity has some relevant consequences. We list here some of them.

(i) The probabilities that appear in the microscopic SR model are epistemic, since they can be interpreted as due to a lack of knowledge about microscopic properties.

(ii) The local and noncontextual picture of the microworld provided by the microscopic SR model is inconsistent with the Bell and the Bell--Kocken--Specker theorems. One it can show that it violates an assumption underlying those theorems, which is usually left implicit. Whenever this assumption is stated explicitly, it proves to be physically problematical \cite{gs96b}, \cite{ga99}, \cite{ga00}, \cite{ga02}, \cite{gp04}, which makes its violation admissible.

(iii) From the viewpoint of the model, QM is a theory that is incomplete in several senses (it does not provide the probabilities ${\mathscr P}_{S}^{t}(F)$ and ${\mathscr P}_{S}^{d}(F)$ and it does not say anything about the distribuition of microscopic properties on physical objects in a given state whenever the objects are not detected). From this viewpoint, a broader theory embodying QM can be envisaged, according to which the quantum probability ${\mathscr P}_{S}(F)$ is considered as a conditional rather than an absolute probability.

(iv) The microscopic properties that appear in the model are hidden parameters, but are not hidden variables in the standard sense. Indeed, it can be proved \cite{ga02}, \cite{gp04}, \cite{ga05} that they are not bound to satisfy in every physical situation the condition introduced by Kocken and Specker as a basic requirement ``for the successful introduction of hidden variables'' \cite{ks67}, \cite{me93}. This explains why microscopic properties are noncontextual.  

(v) The no--registration outcome does not occur because of flaws of the measuring apparatus, but it is determined by the microscopic properties of the physical object. Hence, ${\mathscr P}_{S}^{d}(F)$ may be less than $1$ also in the case of ideal apparatuses.

\section{The quantum machine}
By introducing the entity called \emph{quantum machine} \cite{a86}--\cite{a94b} one can produce a macroscopic model for measurements on a quantum system described by a two--dimensional Hilbert space (\emph{e.g.}, a spin--$\frac{1}{2}$ quantum particle whenever only spin observables are taken into account), which suggests that quantum probabilities can be reinterpreted as epistemic rather than ontologic. 
\begin{figure}
\begin{center}
\includegraphics[height=7.5cm]{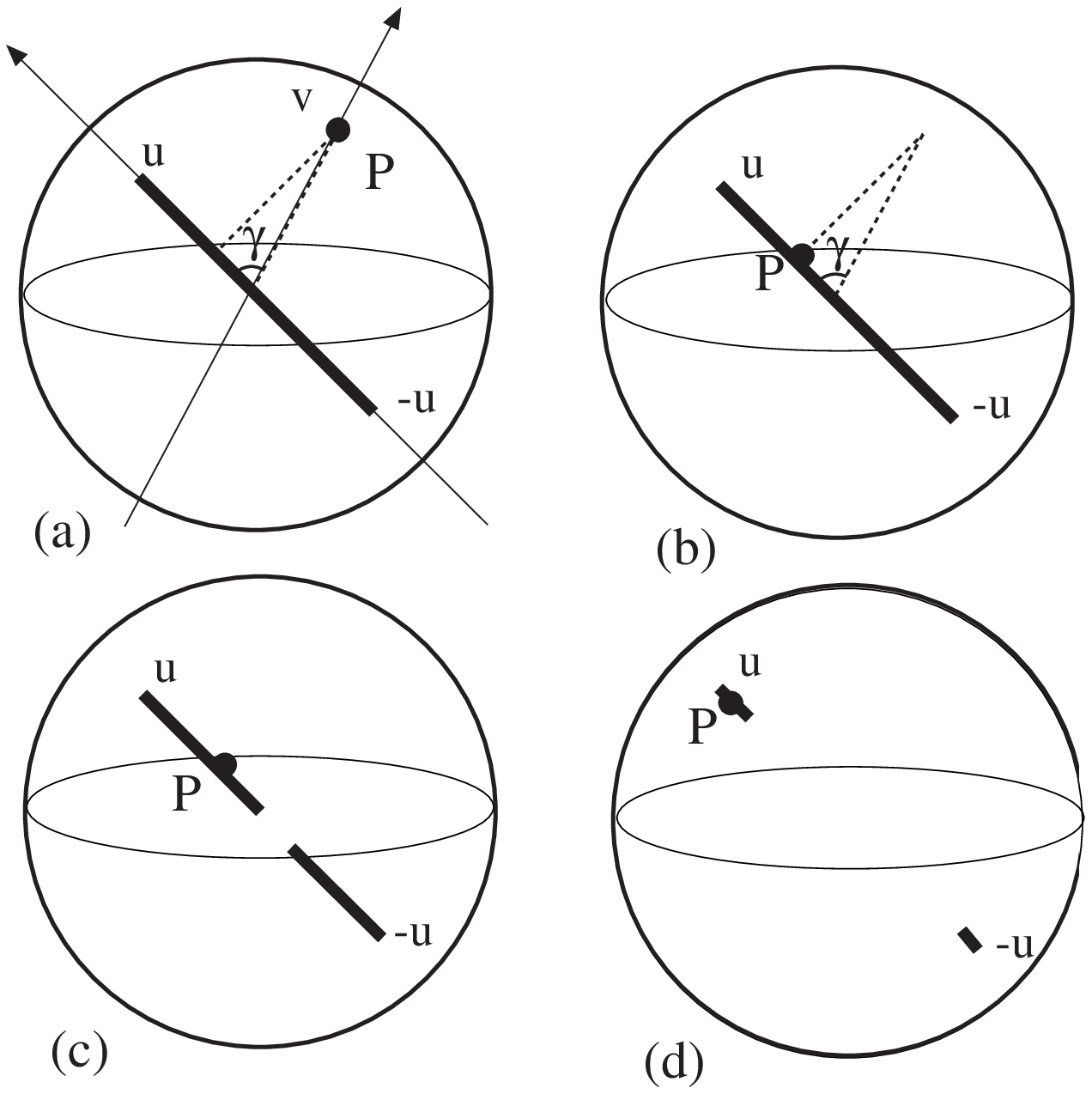}
\end{center}
\textbf{Fig. 2}. \emph{A three dimensional representation of the} quantum machine \emph{as proposed by Aerts.}
\end{figure}

The quantum machine consists of a classical point particle bound to stay on the surface of a spherical ball with radius $1$. Hence each pure state of the machine is represented by a point $P$ of this surface, or, equivalently, by a position vector $\vec v$ belonging to the unitary \emph{Aerts sphere}. Furthermore, each possible experiment connected to the quantum machine can be described as follows. Consider two diametrically opposite points on the Aerts sphere, briefly identified with the (unitary) position vectors $\vec u$ and $-\vec u$ respectively, and install an elastic strip of $2$ units of length, fixed with one of its end--points in $\vec u$ and the other end--point in $-\vec u$ (\textbf{Fig. 2 (a)}). Whenever the experiment is performed, the particle falls from its original place orthogonally onto the elastic, and sticks to it (\textbf{Fig. 2 (b)}). Then, the elastic breaks at some arbitrary point. Consequently the particle, attached to one of the two pieces of the elastic (\textbf{Fig. 2 (c)}), is pulled to one of the two end--points $\vec u$ or $-\vec u$ (\textbf{Fig. 2 (d)}). Now, depending on whether the particle arrives in $\vec u$ or in $-\vec u$, we give the outcome $o_1$ or $o_2$ to the experiment.
\begin{figure}
\begin{center}
\includegraphics[height=5cm]{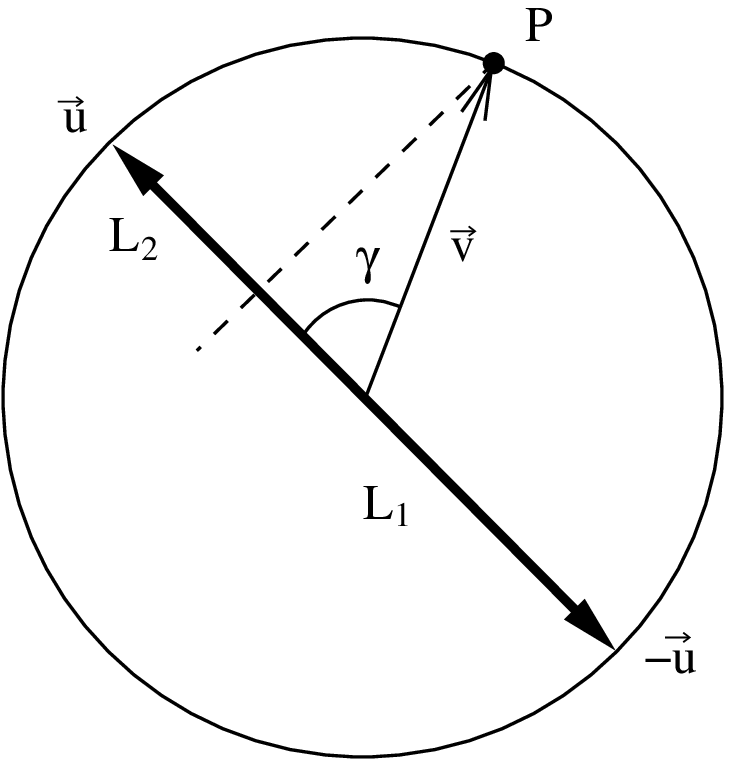}
\end{center}
\textbf{Fig. 3}. \emph{Representation of the experimental process in the plane where it takes place.}
\end{figure}

Let us now calculate the probabilities of the two outcomes. If we demand that the elastic, installed between $\vec u$ and $-\vec u$, can break at any point of this interval with the same probability, the probability $\mu(\vec u, \vec v, o_1)$ that the particle ends up in point $\vec u$, so that the experiment gives outcome $o_1$, when the quantum machine is in the state represented by the vector $\vec v$, is given by the length of the piece of elastic $L_1$ divided by the total length of the elastic (\textbf{Fig. 3}). The probability $\mu(\vec u, \vec v, o_2)$ that the particle ends up in point $-\vec u$, so that the experiment gives outcome $o_2$, when the quantum machine is in the state represented by the vector $\vec v$, is given by the length of the piece of elastic $L_2$ divided by the total length of the elastic. Thus we get
\begin{equation} \label{Aerts1}
\mu(\vec {u},\vec {v},o_{1})=\frac{L_{1}}{2}=\frac{1+\cos {\gamma}}{2}=\cos^{2}{\frac{\gamma}{2}},
\end{equation}
\begin{equation} \label{Aerts2}
\mu(\vec {u},\vec {v},o_{2})=\frac{L_{2}}{2}=\frac{1-\cos {\gamma}}{2}=\sin^{2}{\frac{\gamma}{2}},
\end{equation}
where $\gamma$ is the angle between $\vec u$ and $\vec v$. 

It is well known that the above probabilities coincide with the probabilities that appear in spin measurements on a spin--$\frac{1}{2}$ quantum particle. Indeed, a pure (spin) state of such a particle is represented by the vector
\begin{equation} \label{bloch}
|\psi\rangle=\cos\frac{\theta}{2}e^{-i\frac{\phi}{2}}|+\rangle+\sin\frac{\theta}{2}e^{i\frac{\phi}{2}}|-\rangle
\end{equation}
which is an eigenvector corresponding to the eigenvalue $+\frac{1}{2}\hbar$ of the self--adjoint operator $A=\frac{1}{2}\hbar \vec{\sigma} \cdot \vec{v}$, representing the observable \emph{spin along the direction} $\vec v=\hat{x}\sin\theta\cos\phi+\hat{y}\sin\theta\sin\phi+\hat{z}\cos\theta$. This establishes a correspondence $\omega$ between vectors representing pure states of the spin--$\frac{1}{2}$ quantum particle and points of the surface of a sphere with radius $1$ centered in the origin of $\Re^3$ (\emph{Bloch sphere representation}). This correspondence is one--to--one (up to a phase factor) and obviously induces a bijective mapping of the set of states of the spin--$\frac{1}{2}$ quantum particle on the set of states of the classical point particle considered above. Then, let us consider a measurement of the observable $\mathcal A$ represented by the operator $A=\frac{1}{2}\hbar \vec{\sigma} \cdot \vec{u}$, with $\vec u=\hat{x}\sin\alpha\cos\beta+\hat{y}\sin\alpha\sin\beta+\hat{z}\cos\beta$, on the spin--$\frac{1}{2}$ quantum particle in a state represented by the vector $|\psi\rangle$, and let us denote by $\gamma$ the angle between the two vectors $\vec u$ and $\vec v=\omega(|\psi\rangle)$. It is easy to prove that the probabilities ${\mathscr P}_{\psi}^{A,QM}(+\frac{1}{2}\hbar)$ and ${\mathscr P}_{\psi}^{A,QM}(-\frac{1}{2}\hbar)$ that the measurement yields results $+\frac{1}{2}\hbar$ and $-\frac{1}{2}\hbar$, respectively, are given by
\begin{equation} \label{QM1}
{\mathscr P}_{\psi}^{A,QM}(+\frac{1}{2}\hbar)=\cos^{2}\frac{\gamma}{2},
\end{equation}
\begin{equation} \label{QM2}
{\mathscr P}_{\psi}^{A,QM}(-\frac{1}{2}\hbar)=\sin^{2}\frac{\gamma}{2},
\end{equation}
which coincide with the probabilities $\mu(\vec u, \vec v, o_1)$ and $\mu(\vec u, \vec v, o_2)$ in eqs. (\ref{Aerts1}) and (\ref{Aerts2}), respectively. Hence, our measurement is equivalent to performing an experiment with the quantum machine in the state represented by the vector $\vec v=\omega(|\psi\rangle)$ and the elastic installed between $\vec u$ and $-\vec u$. It follows that the quantum machine provides a macroscopic model for measures of the spin of a spin--$\frac{1}{2}$ quantum particle or, more generally, for any quantum system associated with a two--dimensional complex Hilbert space.

\section{A unified model}
The models described in the previous sections have different features and have been constructed with different aims. The microscopic SR model is a noncontextual general model for measurements on any kind of quantum system, aiming to demonstrate the consistency of the SR interpretation of QM. The quantum machine provides a macroscopic model for measurements on quantum systems described by two dimensional Hilbert spaces, aiming in particular to suggest some enlargements of QM which would allow us to go beyond its present limits, which is classified by the authors as \emph{highly contextual}\cite{a04} (in the sense that the result of a measurement depends also on the measuring apparatus and not only on the state of the particle that is measured). 

There are however some remarkable analogies between the two models. Let us point out some of them. 

First of all, in both models probabilities are epistemic, which follows from the adoption of two ``nonstandard'' hidden variables theories. An measurement on the quantum machine is an example of \emph{hidden measurement}, in the sense that probabilities appear because of lack of knowledge about the specific measurement that is actually performed on the entity (namely, one does not know the specific point in which the elastic breaks), not because of lack of knowledge about states of a quantum object, as in a standard hidden variables theory. The epistemicity of probabilities in the microscopic SR model follows instead both from lack of knowledge about the microstates (as in a standard hidden variables theory) and about the measurement (unknown probability of the $a_0$ outcome in a microstate); this lack of knowledge disappears, however, in a deterministic model, see footnote 1). Hence, the microscopic SR model (which is noncontextual and local) can also be considered a nonstandard hidden variables theory. 

Secondly, both models reproduce quantum probabilities by introducing suitable conditions. Within the quantum machine model only elastic measurements are permitted. Within the microscopic SR model quantum probabilities follow whenever one considers only objects that are actually detected.   

Bearing in mind the above similarities, one may wonder whether a macroscopic version of the microscopic SR model can be constructed which embodies Aerts' quantum machine. At first sight this task seems impossible because of the opposite features of the two models with respect to contextuality. We show in the next sections that the problem can be overcome and construct the desired model.

\subsection{Description of the model for pure states}
Bearing in mind the microscopic SR model discussed in Sec. 2, we modify Aerts' quantum machine as illustrated in \textbf{Fig. 4}. To be precise, let us suppose that the classical point particle that is in the point $P$, hence in the state $S$ represented by the vector $\vec v$ according to Aerts' model, is actually in one of the points on the surface of the second sphere (\emph{detection sphere}), which is identical to Aerts' sphere and tangent to it in the point $P$. Let us install an elastic of $2$ units of length in the direction determined by the vector $\vec v$. Whenever a measurement is performed, the particle falls orthogonally onto the elastic, then the elastic breaks in some arbitrary point and the particle ends up in one of the two extremal points of the elastic. If the particle ends up in the contact point $P$ with Aerts sphere, then we say that the particle is \emph{detected} and, in this case, the experimental process can give one of two outcomes, $o_1$ or $o_2$, with the probabilities predicted by Aerts' model. On the contrary, if the particle ends up in the other extreme, we say that it is \emph{not detected} and, in this case, the outcome $a_0$ is obtained.
\begin{figure}
\begin{center}
\includegraphics[height=6cm]{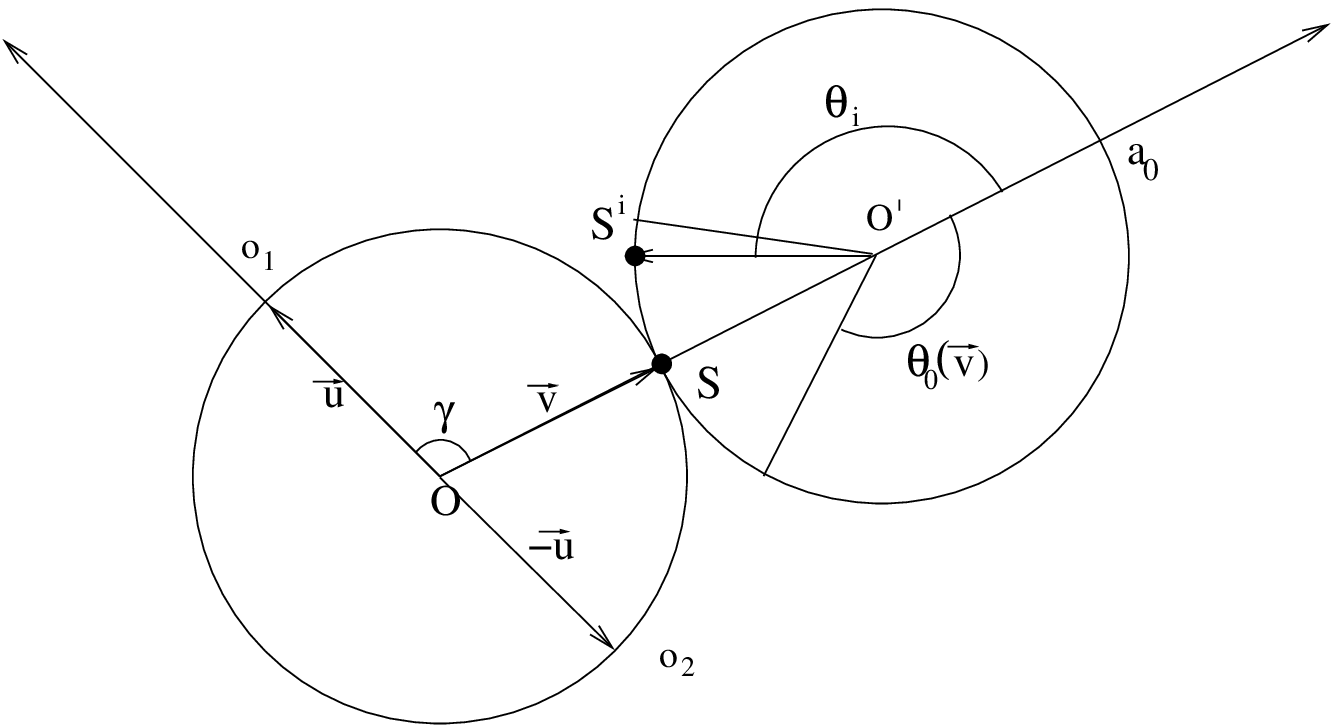}
\end{center}
\textbf{Fig. 4}. \emph{Representation of the detection and the quantum measurement in the plane where they take place.}
\end{figure}

Let us remind that in the microscopic SR model every pure state $S$ of the physical system can be split into microstates $S^{i}$, and let us identify these \emph{hidden} states in our model with the points on the surface of the detection sphere. Furthermore, let us note that the macroscopic property $F$ in Sec. 2 can be identified now either (i) with the pair $(\vec u, o_1)$, or (ii) with the pair $(\vec u,o_2)$. The measurements of both properties are performed in the same way, hence we can assume that the detection probability ${\mathscr P}_{S}^{i,d}(F)$ in eq. (\ref{newprobability-d}) does not depend on $o_1$ and $o_2$ but only on $\vec u$, and write it ${\mathscr P}_{S}^{i,d}(\vec u)$ in our particular case. Therefore the detection probability ${\mathscr P}_{S}^{d}(F)$ will be written  ${\mathscr P}_{S}^{d}(\vec u)$ in our case and eq. (\ref{newprobability-d}) becomes    
\begin{equation}
{\mathscr P}_{S}^{d}({\vec u})=\sum_{i}{\mathscr P}(S^{i}|S) {\mathscr P}_{S}^{i,d}({\vec u}).
\end{equation} 
Let $(\theta_i,\phi_i)$ be the spherical coordinates of the point of the detection sphere which correspond to the microstate $S^i$ when the polar axis is chosen parallel to $\vec v$. Let us reason as in Sec. 3, yet assuming that the probability that the elastic in the detection sphere breaks at some arbitrary point is not the same for every point, but is described by a probability distribution depending on $\gamma$. Hence, we put
\begin{equation}
{\mathscr P}_{S}^{i,d}({\vec u})=p(\gamma,\theta_i),
\end{equation}
whence
\begin{equation}
{\mathscr P}_{S}^{d}({\vec u})=\sum_{i}{\mathscr P}(S^{i}|S)p(\gamma,\theta_i).
\end{equation}
The conditional probabilities ${\mathscr P}(S^{i}|S)$ are not predetermined and it is possible to make assumptions on them. For example, by considering the continuum limit, we can substitute  $S^{i}$ with  $S(\theta,\phi)$ and introduce the following assumptions, based on symmetry arguments.

(i) The conditional probability density is independent of the spherical coordinate $\phi$, hence ${\mathscr P}(S^{i}|S) \longrightarrow {\mathscr P}(S(\theta,\phi)|S)=f(\theta)$.

(ii) Only the hidden states belonging to the surface of a spherical cap $\mathscr C$ centered in $P$ have a conditional probability density $f(\theta)$ different from $0$, and the limit angle $\theta_0$ of the cap depends on the vector $\vec v$, hence we write $\theta_0=\theta_0({\vec v})$. 

Because of the above assumptions, the probability ${\mathscr P}_{S}^{d}({\vec u})$ is given by
\begin{equation}
{\mathscr P}_{S}^{d}({\vec u})=\int_{0}^{2\pi}d\phi\int_{\theta_0(\vec v)}^{{\pi}} f(\theta) p(\gamma,\theta) \sin\theta d\theta.
\end{equation} 

One can make the model more specific by adding suitable assumptions on $p(\gamma,\theta)$ and $f(\theta)$. In any case, if we require that the particle has to be in a definite state, $f(\theta)$ must be such that
\begin{equation}
\int_{{\mathscr C}}f(\theta) d\sigma=\int_{0}^{2\pi}d\phi\int_{\theta_0({\vec v})}^{\pi} f(\theta)  \sin\theta d\theta=1.
\end{equation}

Let us consider now eq. (\ref{macro2}) and the two possibilities (i) $F=(\vec u, o_1)$ and (ii) $F=(\vec u, o_2)$. It is apparent that in case (i) ${\mathscr P}_{S}(F)$ coincides with the probability $\mu(\vec u,\vec v,o_1)$ in eq. (\ref{Aerts1}), while in case (ii) it coincides with $\mu(\vec u,\vec v,o_2)$ in eq. (\ref{Aerts2}). Hence, we get 
\begin{equation} \label{goodform_1}
{\mathscr P}_{S}^{t}\Big ( ({\vec u},o_1) \Big ) ={\mathscr P}_{S}^{d}({\vec u})\mu(\vec u,\vec v,o_1)={\mathscr P}_{S}^{d}(\vec u) \cos^{2}\frac{\gamma}{2},
\end{equation} 
\begin{equation} \label{goodform_2}
{\mathscr P}_{S}^{t} \Big ( ({\vec u},o_2) \Big ) ={\mathscr P}_{S}^{d}(\vec u)\mu(\vec u,\vec v,o_2)={\mathscr P}_{S}^{d}(\vec u)\sin^{2}\frac{\gamma}{2}.
\end{equation}

In order to complete the model from an SR viewpoint, we must still point out two properties $f_{+}$ and $f_{-}$ of the classical point particle which correspond to $(\vec u,o_1)$ and $(\vec u,o_2)$, respectively, and state a criterion for establishing whether $f_{+}$ or $f_{-}$ is possessed by the particle in a given hidden state $S^i$. This can be done as follows. Firstly, partition the spherical cap $\mathscr C$ considered above into a inner spherical cap ${\mathscr C}_{+}$ centered in $P$ and an outer spherical crown ${\mathscr C}_{-}$. Then assume that $\int_{{\mathscr C}_{+}}f(\theta)d\sigma=\cos^{2}\frac{\gamma}{2}$ and  $\int_{{\mathscr C}_{-}}f(\theta)d\sigma=\sin^{2}\frac{\gamma}{2}$. Finally, assume that a particle in a hidden state belonging to ${\mathscr C}_{+}$ (${\mathscr C}_{-}$) produces a breakdown of the elastic in the segment $L_1$ ($L_2$) of \textbf{Fig. 3}, hence outcome $o_1$ ($o_2$). The properties $f_{+}$ and $f_{-}$ are then characterized by the set of hidden states in ${\mathscr C}_{+}$ and ${\mathscr C}_{-}$, respectively, and the factors $\cos^{2}\frac{\gamma}{2}$ and $\sin^{2}\frac{\gamma}{2}$ in eqs. (\ref{goodform_1}) and (\ref{goodform_2}), respectively, are explained in terms of microstates. 

The construction of our unified model is thus concluded. However, this opens a new problem. Indeed, the new model is a macroscopic version of the microscopic SR model, which we classified as noncontextual at the beginning of this section. One may then wonder how it was possible to embody in  it the quantum machine, which provides a model which was classified instead as highly contextual by the authors themselves. The answer to this question is not trivial, and requires a brief preliminary analysis of the concept of contextuality.

According to a standard viewpoint, a physical theory is contextual whenever the value of an observable $\mathcal A$ in a given state of a physical system depends on the set of (compatible) measurements that are simultaneously performed on the system \cite{me93}. We call this kind of contextuality here \emph{contextuality$_1$}, and note that no reference is made in its definition to individual differences between apparatuses measuring $\mathcal A$, which are thus implicitly considered ideal and identical. On the contrary, according to the GB approach the contextuality of the quantum machine follows from the fact that each individual experiment introduces a different set of hidden variables of the measuring apparatus, so that different measurements of the same observable may yield different results \cite{a04}. This provides implicitly a different definition of contextuality, that we call here \emph{contextuality$_2$}, which makes reference to the differences that unavoidably exist between individual apparatuses measuring $\mathcal A$.

Let us come now to the microscopic SR model and to the quantum machine. The former can be classified as noncontextual when contextuality$_1$ is understood (indeed, the result of a measurement depends only on the microscopic properties possessed by the physical object that one is considering, that is, on the microscopic state $S^i$ of the object)\footnote{We remind that the price for noncontextuality$_{1}$ of the microscopic SR model is accepting that the laws of QM cannot be applied to those physical situations that are unaccessible, \emph{in principle}, to empirical control \cite{gs96b}--\cite{ga02}. Such situations actually occur in QM because of the existence of incompatible observables. We stress that this feature of the SR model allows one to avoid a number of paradoxes without conflicting with the theoretical description and the predictions of QM (which refer to detected physical objects only).}. If, on the contrary, contextuality$_2$ is understood and the detection probability ${\mathscr P}_{S}^{i,d}(F)$ in eq. (\ref{newprobability-d}) is interpreted as expressing lack of knowledge on the interaction between the measurement and the physical object, the microscopic SR model can be classified as contextual (but also this kind of contextuality disappears if the SR model is deterministic). Analogously, it is apparent that the measurements on the quantum machine are noncontextual if contextuality$_{1}$ is understood (indeed, measurements with the elastic strip in different directions are never compatible). On the contrary, they are highly contextual, as stated in the GB approach, if contextuality$_{2}$ is understood. Our problem above is thus solved.

It is still interesting to observe that our unified model reduces in some sense the contextuality$_{2}$ of the measurements on the quantum machine because of the final part of our construction above. Indeed, it is apparent that the unknown features of an experiment on the quantum machine (the point in which the elastic breaks), which affect the result of the measurement, are explained within our unified model in terms of the hidden states $S^i$, hence only the contextuality$_{2}$ following from the unknown probability of the $a_0$ outcome is left in the model.

It remains to ``close the circle'' by showing that the above macroscopic model mimics a spin measurement of a spin--$\frac{1}{2}$ quantum particle according to the microscopic SR model. To this end, let us consider the observable $\mathcal A$ represented by the operator $A=\frac{1}{2}\hbar \vec{\sigma} \cdot \vec{u}$ in standard QM (Sec. 3). According to the microscopic SR model, this observable must actually contain in its spectrum, besides the values $+\frac{1}{2}\hbar$ and $-\frac{1}{2}\hbar$, a further value $a_0$ that is considered as the outcome of a measurement when the physical object is not detected, hence it must be substituted by an observable ${\mathcal A}_0$ (we remind that the lack of detection is not interpreted as an inefficiency of the measuring apparatus, but as a consequence of the microscopic properties of the measured object, see Sec. 2).  Let ${\mathscr P}_{\psi}^{A}(+\frac{1}{2}\hbar)$ and  ${\mathscr P}_{\psi}^{A}(-\frac{1}{2}\hbar)$ be the probabilities of finding the outcomes $+\frac{1}{2}\hbar$ and $-\frac{1}{2}\hbar$, respectively, in a measurement of the observable ${\mathcal A}_0$ on a quantum particle in the state represented by the vector $|\psi\rangle$ in eq. (\ref{bloch}). By setting (i) $F=({\mathcal A}_0, \{ +\frac{\hbar}{2} \})$ and (ii)  $F=({\mathcal A}_0, \{ -\frac{\hbar}{2} \})$, these probabilities particularize in two different cases the probability ${\mathscr P}_{S}^{t}(F)$ in eq. (\ref{macro2}). Both in (i) and (ii) the measurement of $F$ is performed by measuring ${\mathcal A}_0$, hence the detection probability ${\mathscr P}_{S}^{d}(F)$ that appears in (\ref{macro2}) is the same in both cases and we briefly denote it by ${\mathscr P}_{\psi}^{d}(A)$. Finally, ${\mathscr P}_{S}(F)$ in eq. (\ref{macro2}) obviously coincides with ${\mathscr P}_{\psi}^{A,QM}(+\frac{1}{2}\hbar)$ (see eq. (\ref{QM1})) in case (i) and with ${\mathscr P}_{\psi}^{A,QM}(-\frac{1}{2}\hbar)$ (see eq. (\ref{QM2})) in case (ii). Thus, we get from eq. (\ref{macro2}),     
\begin{equation} \label{unifiedform_1}
{\mathscr P}_{\psi}^{A}(+\frac{1}{2}\hbar)={\mathscr P}_{\psi}^{d}(A){\mathscr P}_{\psi}^{A,QM}(+\frac{1}{2}\hbar)={\mathscr P}_{\psi}^{d}(A){\cos}^{2}\frac{\gamma}{2},
\end{equation}
\begin{equation} \label{unifiedform_2}
{\mathscr P}_{\psi}^{A}(-\frac{1}{2}\hbar)={\mathscr P}_{\psi}^{d}(A){\mathscr P}_{\psi}^{A,QM}(-\frac{1}{2}\hbar)={\mathscr P}_{\psi}^{d}(A){\sin}^{2}\frac{\gamma}{2}.
\end{equation}        
These equations coincide with eqs. (\ref{goodform_1}) and (\ref{goodform_2}), respectively, if one puts ${\mathscr P}_{\psi}^{d}(A)={\mathscr P}_{S}^{d}({\vec u})$. 

We would like to stress  that our unified model aims to provide a macroscopic analogue of a quantum measurement, but does not claim in any way to explain what actually occurs at a microscopic level. Nevertheless, the unspecified factor ${\mathscr P}_{S}^{d}({\vec u})$ reminds us that, according to the SR interpretation, QM is an incomplete theory which could be embedded, at least in principle, into a broader theory (which is excluded by the standard interpretation). 
 
\subsection{Description of the model for mixed states}
The unified model proposed in the previous section can be generalized to the case of nonpure (mixed) states or \emph{mixtures}. Let us shortly describe this generalization. 

Let us broaden the set of states of the quantum machine by adding the inner points of the Aerts sphere to the points on the surface as possible locations of the classical point particle. Then, let us note that the state $D$ characterized by the vector $\vec w$ such that $|\vec w| < 1$ can be written as a convex combination of the vectors $\vec v=\frac{\vec w}{|\vec w|}$ and $-\vec v=-\frac{\vec w}{|\vec w|}$ representing the pure states $S_1$ and $S_2$, respectively. Indeed, $\vec w=\lambda_1 \vec v+\lambda_2(-\vec v)$, with $\lambda_1=\frac{1+|\vec w|}{2}$ and $\lambda_2=\frac{1-|\vec w|}{2}$ (hence $0 \le \lambda_1,\lambda_2 \le 1$ and $\lambda_1+\lambda_2=1$). If we then perform a measurement of the kind considered in Sec. 3 whenever the quantum machine is in the state $D$ (see \textbf{Fig. 3} with $\vec w$ in place of $\vec v$), the probabilities of obtaining outcomes $o_1$ or $o_2$ are given by
\begin{equation} \label{QM_mixed1}
\mu(\vec u,\vec w,o_1)=\frac{L_1}{2}=\frac{1+|\vec w|\cos\gamma}{2}=\lambda_1 \cos^{2}\frac{\gamma}{2}+\lambda_2 \sin^{2}\frac{\gamma}{2}, 
\end{equation}
and
\begin{equation} \label{QM_mixed2}
\mu(\vec u, \vec w, o_2)=\frac{L_2}{2}=\frac{1-|\vec w|\cos\gamma}{2}=\lambda_1 \sin^{2}\frac{\gamma}{2}+\lambda_2 \cos^{2}\frac{\gamma}{2},
\end{equation}
respectively.

Let us now remind that not only pure states, but also mixed states of spin--$\frac{1}{2}$ quantum particles can be represented on the Bloch sphere. In fact, a mixed state represented in standard QM by the density operator $W=\lambda_1|\psi_1\rangle\langle\psi_1|+\lambda_2|\psi_2\rangle\langle\psi_2|$ (where $|\psi_1\rangle$ and $|\psi_2\rangle$ are normalized and orthogonal vectors in the Hilbert space of the system, $0 \le \lambda_1, \lambda_2 \le 1$, $\lambda_1+\lambda_2=1$) corresponds to the inner point of the Bloch sphere characterized by the vector $\vec w=\lambda_1 \vec v+\lambda_2 (-\vec v)$ (where $\vec v$ and $-\vec v$ are the vectors corresponding to $|\psi_1\rangle$ and $|\psi_2\rangle$, respectively, in the Bloch representation discussed in Sec. 3), with $|\vec w|=|\lambda_1-\lambda_2|$. It is then immediate to see that the probabilities ${\mathscr P}_{W}^{A,QM}(+\frac{1}{2}\hbar)$ and ${\mathscr P}_{W}^{A,QM}(-\frac{1}{2}\hbar)$ predicted by QM for a spin measurement in direction $\vec u$ on a spin--$\frac{1}{2}$ quantum particle in the state represented by $W$ coincide with the probabilities in eqs. (\ref{QM_mixed1}) and (\ref{QM_mixed2}), respectively. Hence, the quantum machine provides a macroscopic model for this kind of measurements also in the case of mixed states. It must be stressed, however, that, according to Aerts, $D$ is not interpreted as a mixed state of the quantum machine, which is highly relevant in Aerts' perspective \cite{a99b}.

\begin{figure}
\begin{center}
\includegraphics[height=7cm]{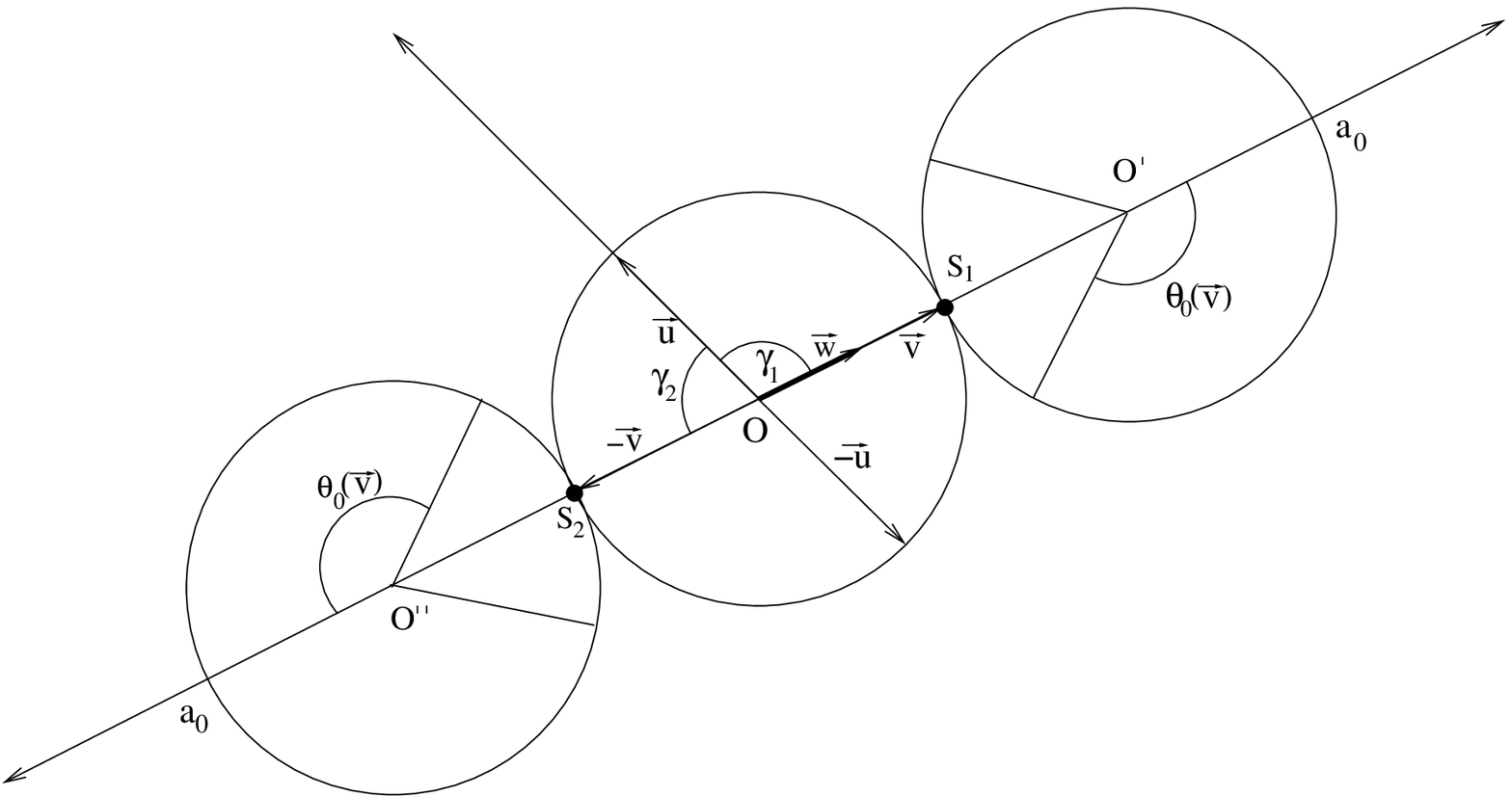}
\end{center}
\textbf{Fig. 5}. \emph{Representation of the detection and the quantum measurement for a mixed state in the plane where they take place.}
\end{figure}      
Let us evaluate the probabilities ${\mathscr P}_{D}^{t}\Big (({\vec u},o_1) \Big )$ and ${\mathscr P}_{D}^{t}\Big ( ({\vec u},o_2) \Big )$ of finding the outcomes $o_1$ and $o_2$, respectively, in a measurement of the kind described in Sec. 4.1 whenever the quantum machine is in the state $D$ within our unified model. If we consider (contrary to Aerts) the state $D$ as a mixture of the states $S_1$ and $S_2$, the coefficients $\lambda_1$ and $\lambda_2$ can be interpreted as the probabilities that the quantum machine in the state $D$ is actually in the state $S_1$ or in the state $S_2$, respectively. Hence, we get
\begin{equation}  \label{SR_mixed1}
{\mathscr P}_{D}^{t} \Big (({\vec u},o_1) \Big )=\lambda_1{\mathscr P}_{S_1}^{t}\Big ( ({\vec u},o_1)\Big )+\lambda_2 {\mathscr P}_{S_2}^{t}\Big (({\vec u},o_1) \Big ), 
\end{equation}
\begin{equation} \label{SR_mixed2}
{\mathscr P}_{D}^{t} \Big (({\vec u},o_2) \Big )=\lambda_1{\mathscr P}_{S_1}^{t}\Big ( ({\vec u},o_2) \Big )+\lambda_2 {\mathscr P}_{S_2}^{t}\Big ( ({\vec u},o_2) \Big ). 
\end{equation}
The probabilities ${\mathscr P}_{S_1}^{t} \Big (({\vec u},o_1)\Big )$, ${\mathscr P}_{S_2}^{t}\Big (({\vec u},o_2)\Big )$, etc., can be calculated by using eqs. (\ref{goodform_1}) and (\ref{goodform_2}). One gets, with $\gamma_1$ and $\gamma_2$ as in \textbf{Fig. 5}, ${\mathscr P}_{S_1}^{t} \Big (({\vec u}, o_1) \Big )={\mathscr P}_{S_1}^{d}(\vec u)  \cos^{2}\frac{\gamma_1}{2}$, ${\mathscr P}_{S_2}^{t} \Big (({\vec u}, o_1) \Big )={\mathscr P}_{S_2}^{d}(\vec u)  \cos^{2}\frac{\gamma_2}{2}$, etc. Since, now, $S_1$ and $S_2$ are represented by the opposite vectors $\vec v$ and $-\vec v$, respectively, and $\gamma_2=\pi-\gamma_1$, the symmetries of the particular physical system at issue suggest to assume that ${\mathscr P}_{S_1}^{d}(\vec u)={\mathscr P}_{S_2}^{d}(\vec u)$. By setting ${\mathscr P}_{S_1}^{d}(\vec u)={\mathscr P}_{S_2}^{d}(\vec u)={\mathscr P}_{D}^{d}(\vec u)$ and $\gamma_1=\gamma$, we get
\begin{equation} \label{SR_mixed1f}
{\mathscr P}_{D}^{t}\Big (({\vec u}, o_1)\Big )={\mathscr P}_{D}^{d}(\vec u)(\lambda_1 \cos^{2}\frac{\gamma}{2}+\lambda_2\sin^{2}\frac{\gamma}{2}),
\end{equation}
\begin{equation} \label{SR_mixed2f}
{\mathscr P}_{D}^{t} \Big (({\vec u}, o_2)\Big )={\mathscr P}_{D}^{d}(\vec u)(\lambda_1 \sin^{2}\frac{\gamma}{2}+\lambda_2\cos^{2}\frac{\gamma}{2}).
\end{equation}
Proceeding as in Sec. 4.1, the above probabilities can then be identified with the probabilities ${\mathscr P}_{W}^{A}(+\frac{1}{2}\hbar)$ and ${\mathscr P}_{W}^{A}(-\frac{1}{2}\hbar)$, respectively, that a measurement of the observable ${\mathscr A}_0$ on a spin--$\frac{1}{2}$ quantum particle in the state represented by $W$ yields outcome $+\frac{1}{2}\hbar$ and $-\frac{1}{2}\hbar$, respectively, according to the microscopic SR model. 

Our unified model has thus been generalized to the case of mixtures, as desired. It must be stressed, however, that this has been done at the expense of betraying Aerts' original idea of not considering the state $D$ of the quantum machine as a mixture.

\section{Conclusions}
The construction in Sec. 4 shows that Aerts' quantum machine can be used as a basis for producing a more complex model for quantum measurements on spin--$\frac{1}{2}$ particles. The new model constitutes a macroscopic version of the microscopic model for quantum measurements introduced within the SR interpretation, hence it establishes a first formal link between the GB approach and the SR interpretation of QM. Moreover, some relevant differences between the two approaches, which seemingly make them incompatible, are bypassed in the model. This suggests that they can be bypassed in general by using similar procedures, even if some difficulties could arise when considering models for quantum systems described by Hilbert spaces whose dimension is greater that 2. In any case, the model presented in this paper may serve as an intuitive basis for the attempt at linking together the GB approach and the SR intepretation, aiming to construct a broader theory going beyond the present limits of QM.

\end{document}